\documentclass[amsmath,amssymb,aps,aps,showkeys,showpacs,dvips]{revtex4}

\usepackage[dvips]{graphicx}
\usepackage[T1]{fontenc}
\usepackage[latin2]{inputenc}
\usepackage[english]{babel}
\usepackage{latexsym}
\usepackage{bbm}

\newcounter{cdef}
\newcommand{\definition}{{\stepcounter{cdef}\bf Definition \thecdef{}. }}
\newcounter{cpro}
\newcommand{\proposition}{{\stepcounter{cpro}\bf Proposition \thecpro{}. }}
\newcounter{cexa}
\newcommand{\example}{{\stepcounter{cexa}\bf Example \thecexa{}. }}
\begin{document}
\title{Quantum Iterated Function Systems}
\author{Artur {\L}ozi{\'n}ski}\email{lozinski@if.uj.edu.pl}
\affiliation{Instytut Fizyki im. Smoluchowskiego, Uniwersytet
Jagiello{\'n}ski, ul. Reymonta 4, 30-059 Krak{\'o}w, Poland}

\author{Karol {\.Z}yczkowski}\email{karol@cft.edu.pl}
\affiliation{Centrum Fizyki Teoretycznej PAN, Al. Lotnik{\'o}w 32/46, 02-668
Warszawa, Poland\\
and Instytut Fizyki UJ, Krak{\'o}w, Poland}

\author{Wojciech S{\l}omczy{\'n}ski}\email{slomczyn@im.uj.edu.pl}
\affiliation{Instytut Matematyki, Uniwersytet
Jagiello{\'n}ski, ul. Reymonta 4, 30-059 Krak{\'o}w, Poland}

\begin{abstract}
Iterated functions system (IFS) is defined by specifying a set of
functions in a classical phase space, which act randomly on an initial
point. In an analogous way, we define a quantum iterated functions
system (QIFS), where functions act randomly with prescribed
probabilities in the Hilbert space. In a more general setting a QIFS
consists of completely positive maps acting in the space of density
operators. This formalism is designed to describe certain problems of
nonunitary quantum dynamics. We present exemplary classical IFSs, the
invariant measure of which exhibits fractal structure, and study
properties of the corresponding QIFSs and their invariant states.
\end{abstract}

\date{\today}
\maketitle

\section{Introduction}

An {\sl iterated function system} (IFS) may be considered as a
generalization of a classical dynamical system, which permits a certain
degree of stochasticity. It is defined by a set of $k$ functions $f_i :
\Omega \to \Omega, \thinspace i=1,\dots,k$, which represent discrete dynamical
systems in the classical phase space $\Omega$. The functions $f_i$ act
randomly with given place-dependent probabilities $p_i : \Omega \to
[0,1], i=1,\dots,k, \sum_{i=1}^{k}p_{i}=1$ \cite{Ba}. They characterize
the likelihood of choosing a particular map at each step of the time
evolution of the system.

There exist different ways of investigating such random systems. Having
defined an IFS, one may ask, how is an initial point $x_0\in \Omega$
transformed by the random system. In a more general approach, one may
pose a question that how does a probability measure $\mu$ on $\Omega$
change under the action of the Markov operator $P$ associated with the IFS.
If the phase space $\Omega$ is compact, the functions $f_i$ are strongly
contracting, and the probabilities $p_i$ are H\"{o}lder continuous and
positive (i.e. $p_i > 0$), then there exists a unique invariant measure
$\mu_{*}$ of $P$ -- see for instance \cite{Ba,K81,BDEG88}, and references
therein.

For a large class of IFSs, the invariant measure $\mu_*$ has a fractal
structure. Such IFSs may be used to generate fractal sets in the space
$\Omega$. In particular, iterated function systems leading to well-known
fractal sets, such as the Cantor set or the Sierpi{\'n}ski gasket, 
can be found in Ref.~\cite{Ba}. These intriguing properties of IFSs allowed
one to apply them for image compression, processing, and encoding
\cite{Ba,BE88,BaBY}.

Iterated function systems can also be used to describe several physical
problems, where deterministic dynamics is combined with the random
choice of interaction. In particular, IFSs belong to a larger class of
random systems studied in Ref.~\cite{YOC91,PSV95}. Such a composition of
deterministic and stochastic behavior is important in numerous fields of
science, since very often an investigated dynamical system is subjected
to an external noise.

Nondeterministic dynamics may also be relevant from the point of view
of quantum mechanics. Although unitary time evolution of a closed
quantum system is purely deterministic, the problem  changes if one tries
to take into account processes of quantum measurement or a possible
coupling with a classical system. In the approach of Event Enhanced
Quantum Theory (EEQT) developed by Blanchard and Jadczyk \cite{BJ95},
the quantum time evolution is piecewise deterministic and in certain
cases may be put into the framework of iterated function systems
\cite{BJO98,BJO99}.  While some recent investigations in this area
concentrate mostly on IFS's acting in the space of pure states
\cite{JO02}, we advocate a more general setup, in which IFS's act in the
space of mixed quantum states.

The main objective of this paper is to propose a general definition of {\sl
quantum iterated function system} (QIFSs). Formally, it suffices to
consider the standard definition of IFS and to take for $\Omega$ an
$N$-dimensional Hilbert space $ {\cal H}_N$. Instead of functions $f_i,
\thinspace i=1,\dots,k$, representing classical maps, one should use
linear functions $F_i: {\cal H}_N \to {\cal H}_N$, which represent the
corresponding quantum maps. Alternatively, one may consider the space
${\cal M}_N$ of density matrices of size $N$ and construct an iterated
function system out of $k$ positive maps $G_i:{\cal M}_N \to {\cal
M}_N$.  The QIFSs defined in this way can be used to describe processes
of quantum measurements, decoherence, and dissipation. Moreover, QIFSs
offer an attractive field of research on the semiclassical limit of
quantum random systems. In particular, it is interesting to explore
quantum analogues of classical IFSs, which lead to fractal invariant
measures, and to investigate, how do quantum effects smear fractal
structures out.

This paper is organized as follows. In the following section we recall the
definition and basic properties of the classical IFSs, and discuss several
examples. In Sec.~III we propose the definition of QIFSs, investigate
their properties, and relate them to the notion of quantum channels and
complete positive maps used in the theory of quantum dynamical semigroups.
The quantum--classical correspondence is a subject of Sec.~IV, in which
we compare dynamics of exemplary IFSs and the related QIFSs. 
Concluding remarks are presented in Sec.~V.

\section{Classical iterated function systems}

Consider a compact metric space $\Omega$ and $k$ functions
$f_i:\Omega\to \Omega$, where $i=1,\dots,k$. Let us specify $k$
probability functions $p_i:\Omega \to [0,1]$ such that for each point
$x\in \Omega$ the condition $\sum_{i=1}^k p_i(x)=1$ is fulfilled. Then
the functions $f_i$ may be regarded as classical maps, which act
randomly with probabilities $p_i$. The set ${\cal F}_{\rm Cl}:=\{
\Omega, f_i, p_i : i=1,\dots,k \}$ is called an {\sl iterated
function system} (IFS).

Let ${\cal M}(\Omega)$ denotes the space of all probability measures on
$\Omega$. The IFS ${\cal F}_{\rm Cl}$ generates the following {\sl
Markov operator} $P$ acting on ${\cal M}(\Omega)$
\begin{equation}
(P\mu)(B) = \sum_{i=1}^{k} \int_{f_i^{-1}(B)} p_i(x) d\mu(x) 
\text{ ,}
\label{IFS1}
\end{equation}
where $B$ is a measurable subset of $\Omega$ and a measure $\mu$ belongs
to ${\cal M}(\Omega)$.  This operator represents the corresponding
Markov stochastic process defined on the code space consisting of
infinite sequences built out of $k$ letters which label each map $f_i$.
On the other hand, $P$ describes the {\sl evolution of probability
measures} under the action of ${\cal F}_{\rm Cl}$.

Consider an IFS defined on an interval in ${\mathbb R}$ and consisting
of invertible $C^1$ maps $\{f_i:i=1,\dots,k\}$. This IFS generates the
associated Markov operator $P$ on the space of densities \cite{LM94},
which describes one step evolution of a classical density $\gamma$
\begin{equation}
P[\gamma](x) = \sum_{i} p_i\bigl(f_i^{-1}(x)\bigr)\gamma\bigl(f_i^{-1}(x)\bigr) 
\left| \frac{d f_i^{-1}(x)}{dx} \right|
\text{ ,}
\label{IFS6}
\end{equation}
where for $x \in \Omega$ the sum goes over $ i=1,\dots , k$, such that
$x \in f_{i}(\Omega)$.

Let $d(x,y)$ denotes the distance between two points $x$ and $y$ in the
metric space $\Omega$. An IFS ${\cal F}_{\rm Cl}$ is called {\sl
hyperbolic}, if it fulfills the following conditions for all
$i=1,...,k$:

(i) $f_i$ are Lipschitz functions with the Lipschitz constants $L_i <
1$, i.e., they satisfy the contraction condition $d(f_i(x),f_i(y)) \le
L_i d(x,y)$ for all $x,y \in \Omega$;

(ii) the probabilities $p_i$ are H\"{o}lder continuous, i.e.,
 they fulfill the condition $\left|  p_{i}\left(  x\right)  -p_{i}
\left(y\right)  \right|  \leq K_{i}d\left(  x,y\right)  ^{\alpha}$ for some
$\alpha\in\left(  0,1\right]  $, $K_{i}\in\mathbb{R}^{+}$
for all $x, y \in \Omega$;

(iii) all probabilities are positive, i.e.,
$p_i(x) > 0$ for any $x\in \Omega$.

The Markov operator $P$ associated with a hyperbolic IFS has a unique
{\sl invariant probability measure} $\mu_*$ satisfying the equation $P
\mu_* = \mu_*$. This measure is {\sl attractive}, i.e., $P^n{\mu}$
converges weakly to $\mu_*$ for every $\mu \in {\cal M}(\Omega)$ as $n
\to \infty$. In other words, $\int_{\Omega} u ~d P^n{\mu}$ tends to
$\int_{\Omega} u ~d{\mu_*}$ for every continuous function $u : \Omega
\to {\mathbb R}$. Let us mention that the hyperbolicity conditions
(i)-(iii) are not necessary to assure the existence of a unique
invariant probability measure - some other, less restrictive, sufficient
assumptions were analyzed in
Refs.~\cite{K81,BDEG88,LY94,LasMyj99,FanLau99,Ste00,Sza00}.

Observe that in the above case, in order to obtain the exact value of an
integral $\int_{\Omega} u ~d{\mu_*}$, it is sufficient to find the limit
of the sequence $\int_{\Omega} u ~d(P^n{\mu})$ for an arbitrary initial
measure $\mu$. This method of computing integrals over the invariant
measure $\mu_*$ is purely  {\sl deterministic} \cite{Ba}. Sometimes it
is possible to perform the integration over the invariant measure
analytically, even though $\mu_*$ displays fractal properties
\cite{SKZ00}. Alternatively a {\sl random iterated algorithm} may be
employed by generating a random sequence $x_j\in \Omega$ by the IFS,
$j=0,1,\dots$, which originates from an arbitrary initial point $x_0$.
Due to the ergodic theorem for IFSs \cite{K81,E87,IG90}, the mean value
$(1/n)\sum_{j=0}^{n-1} u(x_j)$ converges with probability one in the limit $n
\to \infty$ to the desired integral $\int_{\Omega} u ~d{\mu_*}$ for a
large class of $u$.

If probabilities $p_i$ are constant we say that an IFS is {\sl of the
first kind}. Such IFSs are often studied in the mathematical literature
(see Ref.~\cite{Ba} and references therein). Moreover they have also
some applications in physics. For example, they were used to construct
multifractal energy spectra of certain quantum systems \cite{GM94}, and
to investigate second order phase transitions \cite{Ra95}. On the other
hand, IFSs with place-dependent probabilities can be associated with
some classical and quantum dynamical systems
\cite{BDEG88,IG90,GB95,KSZ97,Sl97,OPSZ00,Slo02}. In analogy with the
position-dependent gauge transformations such IFSs may be called {\sl
iterated function systems of the second kind} \cite{SKZ00}.

If $\Omega$ is a compact subset of ${\mathbb R}^n$, while $d_E(x,y)$
represents the Euclidean distance, or $\Omega$ is a compact manifold
(e.g. sphere $S^2$ or torus $T^n$) equipped with the natural
(Riemannian) distance $d_R$, then an IFS will be called {\sl classical}. For
concreteness we provide below some examples of classical IFSs. The first
example demonstrates that even simple linear maps $f_i$ may lead to a
nontrivial structure of the invariant measure.

\medskip

\example  $\Omega=[0,1]$, $k=2$, $p_1=p_2=1/2$ and two
affine transformations are given by $f_1(x)=x/3$ and 
$f_2(x)=x/3+2/3$
for $x \in \Omega$. Since both functions are continuous contractions
with Lipschitz constants $L_1=L_2=1/3<1$, this IFS is hyperbolic. Thus,
there exists a unique attracting invariant measure $\mu_*$. It is easy
to show \cite{Ba} that $\mu_*$ is concentrated uniformly on the Cantor
set of the fractal dimension $d=\ln 2/\ln 3$.

\medskip

The next example presents an IFS of the second kind.

\medskip

\example  As before, $\Omega=[0,1]$, $k=2$, $f_1(x)=x/3$,
and $f_2(x)=x/3+2/3$ for $x \in \Omega$. The probabilities are now place
dependent, $p_1(x)=x$ and $p_2(x)=1-x$. Although this IFS is not
hyperbolic (condition (iii) is not fulfilled), a unique
invariant measure $\mu_*$ still exists. It is also concentrated on the Cantor set,
but now in a non-uniform way \cite{SKZ00}. The measure $\mu_*$ displays in this
case multifractal properties, since its generalized dimension depends on the
R{\'e}nyi parameter.

\medskip

\example $\Omega=[0,1]\times [0,1] \subset {\mathbb
R}^2$, $k=4$, $p_1=p_2=p_3=p_4=1/4$. Four affine transformations are
given by
\begin{eqnarray} f_{1}
	\left( \begin{array}[c]{c} x \\ y \end{array} \right)
	=\left( \begin{array}[c]{cc} 1/3 & 0\\ 0 & 1 \end{array}\right)
	\left( \begin{array}[c]{c} x \\ y \end{array} \right),
	\quad \quad
	f_{2} \left( \begin{array}[c]{c} x \\ y \end{array} \right) =\left(
	\begin{array}[c]{cc}1/3 & 0\\ 0 & 1 \end{array} \right) \left(
	\begin{array}[c]{c} x \\ y \end{array} \right) + \left( \begin{array}
	[c]{c} 2/3 \\ 0 \end{array} \right), \nonumber \\
	f_{3} \left( \begin{array}[c]{c} x \\ y \end{array} \right) =
	\left( \begin{array}[c]{cc}1 & 0\\ 0 & 1/3 \end{array} \right)
	\left( \begin{array}[c]{c}x \\ y \end{array} \right) ,
	\quad \quad
	f_{4} \left( \begin{array}[c]{c} x \\ y \end{array} \right)
	=\left( \begin{array}[c]{cc} 1 & 0\\ 0 & 1/3 \end{array} \right)
	\left( \begin{array}[c]{c} x \\ y \end{array} \right) + \left(
	\begin{array}[c]{c} 0 \\ 2/3 \end{array} \right) .
\label{clascant}
\end{eqnarray}

\begin{figure}[hbt] \begin{center}
\includegraphics[width=.9\textwidth]{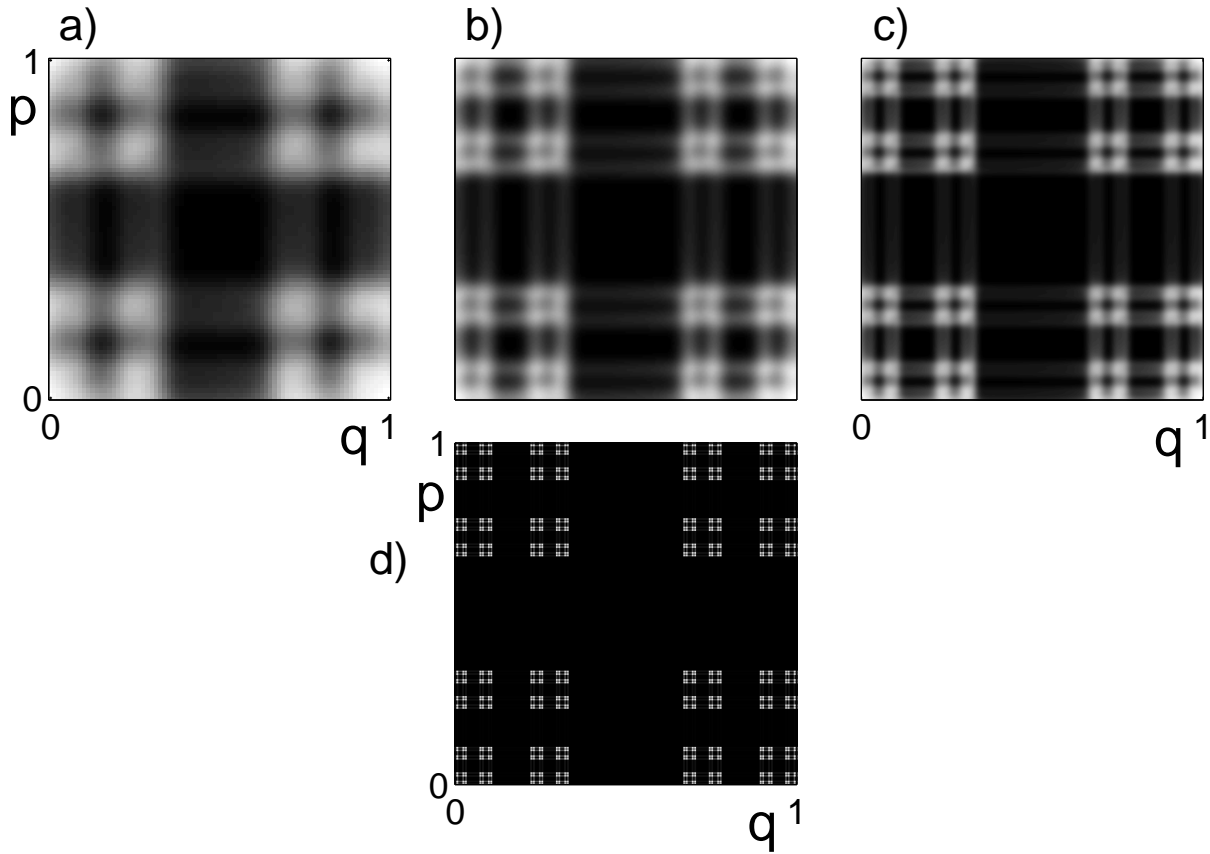}
\end{center}
\caption{"Tartan-like" invariant density of the QIFS defined in
Example~14 for (a)$N=3^4$, (b) $N=3^5$, and (c)~$N=3^6$--dimensional
Hilbert space, shown in the generalized Husimi representation.
Invariant measure of the corresponding classical IFS on the torus
Eq.~(\ref{clascant}) occupies a fractal set (d).}
\label{fig1}
\end{figure}

Also, this IFS is not hyperbolic, since the transformations $f_i$ are not
globally contracting, the former two contract along $x$--axis, while the
latter two along  $y$ axis only. An invariant measure $\mu _{*}$ for
this IFS is presented in Fig.~\ref{fig1}d. The support of $\mu _{*}$ is
the Cartesian product of two Cantor sets. Thus, its fractal dimension is
$d=2\ln 2/\ln 3$.

\medskip

\example Let $\Omega=S^2$. Take $k=2$, 
$p_1=p_2=1/2$, and choose $f_1$ to be the rotation along $z$--axis by angle
$\chi_1$ (later referred to as $R_z(\chi_1)$). In the standard spherical
coordinates, $f_1(\theta,\phi)=(\theta,\phi+\chi_1)$. The second function
$f_2$ is a rotation by angle $\chi_2$ along an axis inclined by angle
$\beta$ with respect to $z$--axis. Since both classical maps are
isometries this IFS is by no means hyperbolic. The properties of the
Markov operator depend on the angle $\beta$, and the commensurability of
the angles $\chi_i$. However, the Lebesgue measure on the sphere is always an
invariant measure for this IFS.

\medskip

\example  $\Omega=[0,1]$, $k=2$, $p_1=p_2=1/2$, $f_1(x)= 2x$ for
$x<1/2$, and $f_1(x)=2(1-x)$ for $x\ge 1/2$ ({\sl tent map}); $f_2(x)=
2x$ for $x<1/2$ and $f_2(x)=2x-1$ for $x\ge 1/2$ ({\sl Bernoulli map}).
Both classical maps are expanding (and chaotic), thus the IFS is not
hyperbolic. The Lebesgue measure in $[0,1]$ is an invariant measure
$\mu_*$ for this IFS.

\medskip
\section{Quantum iterated function systems}

\subsection{Pure states QIFSs}

To describe a quantum dynamical system we consider a complex Hilbert
space $\cal H$. When the corresponding classical phase space $\Omega$ is
compact, the Hilbert space ${\cal H}_N$ is finite dimensional and its dimension 
$N$ is inversely proportional to the Planck constant $\hbar$ measured in the 
units of the volume of $\Omega$. Analyzing quantum systems, $N$ is usually treated 
as a free parameter, and the semiclassical limit is studied by letting $N \to 
\infty$.

A quantum state can be described by an element $|\psi\rangle$ of 
${\cal H}_N$ normalized according to $\langle \psi| \psi \rangle = 1$. 
Since for any phase~$\alpha$ the element 
$|\psi'\rangle = e^{i\alpha}|\psi\rangle$
describes the same physical state as $|\psi\rangle$, we identify them, and so
the space of all pure states ${\cal P}_N$ has $2N-2$ real dimensions. From the
topological point of view, it can be represented as the complex projective
space ${\mathbb C}P^{N-1}$ equipped with the Fubini--Study (FS) metric
given by
\begin{equation}
	D_{FS}(|\phi\rangle, |\psi\rangle)= \arccos
	|\langle \phi| \psi\rangle| \text{ .}
\label{Fubstud}
\end{equation}
It varies from zero for $|\phi\rangle=|\psi\rangle$ to $\pi/2$ for any two
orthogonal states. In the simplest case of a two-dimensional Hilbert space
${\cal H}_2$ the space of pure states ${\cal P}_2$ reduces to the Bloch sphere,
${\mathbb C} P^{1} \simeq S^2$, and the FS distance between any two quantum
states equals to the natural (Riemannian) distance between the corresponding
points on the sphere of radius $1/2$. 

\medskip

\definition To define a {\sl (pure states) quantum iterated function
system} ({\sl QIFS}) it is sufficient to use the general definition of
IFS given in Sect.~II, taking for $\Omega$ the space ${\cal P}_N$.  We
specify two sets of $k$ linear invertible operators: 

\begin{itemize}
\item  $V_{i}:{\cal H}_{N}\rightarrow {\cal H}_{N}$ ($i=1,\ldots ,k$),
which generates maps $F_{i}:{\cal P}_{N}\rightarrow {\cal P}_{N}$
($i=1,\ldots ,k$) by
\begin{equation}
F_{i}\left( |\phi\rangle \right) :=\frac{V_{i}\left(|\phi\rangle\right) }{\left\| V_{i}\left(
|\phi\rangle \right) \right\| }  \text{.}
\label{maps}
\end{equation}

\item  $W_{i}:{\cal H}_{N}\rightarrow {\cal H}_{N}$ ($i=1,\ldots ,k$),
forming an operational resolution of identity,
$\sum_{i=1}^{k}W_{i}^{\dagger }W_{i}={\mathbbm 1}$, which generates
probabilities $p_{i}:{\cal P}_{N}\rightarrow \left[ 0,1\right] $ ($
i=1,\ldots ,k$) by
\begin{equation}
p_{i}\left( |\phi\rangle \right) :=\left\| W_{i}\left( |\phi\rangle \right) \right\| ^{2}
\label{probabilities}
\end{equation}
\end{itemize}
for any $|\phi\rangle \in {\cal P}_N.$

Clearly, for any $|\phi\rangle \in {\cal P}_N$ the normalization condition
 $\sum_{i=1}^{k}p_{i}( |\phi\rangle )=1$ is fulfilled. In this 
situation a QIFS may be defined as a set 
\begin{equation}
{\cal F}_{N}=\{{\cal P}_{N} ;\ \ F_{i}:{\cal P}_{N}\rightarrow {\cal P}_{N};  \ \ 
p_{i}:{\cal P}_{N}\rightarrow \left[ 0,1\right] :i=1,...,k\}
\text{ .}
\label{QIFSa}
\end{equation}

\medskip

Such a QIFS may be realized by choosing an initial state $|\phi_0\rangle
\in {\cal P}_N$ and generating randomly a sequence of pure states
$(|\phi_j\rangle)_{j \in {\mathbb N}}$. The state $|\phi_0\rangle$ is transformed into
$|\phi_1\rangle =F_i(|\phi_0\rangle)$ with probability $p_i(|\phi_0\rangle)$, 
later $\left|  \phi_{1}\right\rangle $ is mapped into $\left|  \phi_{2}\right\rangle
=F_{j}\left(  \left|  \phi_{1}\right\rangle \right)  $ with probability
$p_{j}\left(  \left|  \phi_{1}\right\rangle \right)$, and so on. If we choose 
$W_{i}=\sqrt{p_{i}}\,\mathbbm{1}$, then the probabilities are constant: 
$p_{i}\left(|\phi\rangle \right) = p_{i}$ for $i=1,\ldots ,k$. An 
arbitrary QIFS ${\cal F}_N$ determines by formula (\ref{IFS1}) the operator $P$ 
acting on probability measures on ${\cal P}_N$.

\medskip

Such defined QIFS ${\cal F}_N$ cannot be hyperbolic, since the quantum
map $F_i$ are not contractions with respect to the Fubini--Study
distance in ${cal P}_N$.

\medskip

\example  $\Omega={\cal P}_N \simeq {\mathbb C}P^{N-1}$, $k=2$, $p_1=p_2=1/2$,
$F_1(|\psi\rangle)=U_1(|\psi\rangle)$ and
$F_2(|\psi\rangle)=U_2(|\psi\rangle)$, where the operators $U_i$ ($i=1,2$) are
unitary. In this case both quantum maps are isometries.
Thus the natural Riemannian (Fubini-Study) measure in ${\cal P}_N$ is
invariant, but as we shall see in the next section, its uniqueness depends on
the choice of $U_1$ and $U_2$.

\subsection{Mixed states QIFSs}

Mixed states are described by $N-$dimensional density operators $\rho$, i.e., 
positive Hermitian operators acting in $\cal{H}_N$ with trace normalized to 
unity, $\rho =\rho ^{\dagger }$, $\rho \geq 0$ and $\rm tr\rho =1$. They 
may be represented (in a non unique way) as a convex combination of projectors. 
We shall denote the space of density operators by ${\cal M}_N$. 

\medskip

\definition Now we can formulate the general definition of a QIFS as a
set 
\begin{equation} {\cal F}_N:= \{ {\cal M}_N, \; G_i:{\cal M}_N \to {\cal M}_N, 
\; p_i: {\cal M}_N \to [0,1]; \; i=1,...,k \}
\text{ ,}
\label{QIFSb} 
\end{equation} 
where the maps $G_i$, $i=1,\dots,k$ transform density operators into density
operators, and for every density operator $\rho \in {\cal M}_N$ the 
probabilities are normalized, i.e., $\sum_{i=1}^k p_i(\rho)=1$.

\medskip

The above definition of QIFS is more general than the previous one, since in
particular $G_i$ and $p_i$ may be defined by 
\begin{equation}
G_{i}\left( \rho \right) =\frac{V_{i}\rho V_{i}^{\dagger }}
{{\rm tr} \left( V_{i}\rho V_{i}^{\dagger }\right) }
\label{mapsmix}
\end{equation}
and
\begin{equation}
p_{i}\left( \rho \right) = {\rm tr} 
\left( W_{i}\rho W_{i}^{\dagger }\right)
\label{probabilitiesmix}
\end{equation}
for $i=1,\dots,k$ and $\rho \in {\cal M}_N$, where the linear maps $V_i$ and 
$W_i$ are as in Definition~1. Thus, each QIFS on ${\cal P}_N$ can be extended to 
a QIFS on ${\cal M}_N$. Note that in this case $ p_{i}\left( \rho \right) = 
\mathop{\rm tr} ( W_{i}^{\dagger}W_{i}\rho )$. Hence, we can alternatively define 
the probabilities by $p_i(\rho)= \mathop{\rm tr} \left( {\cal L}_{i}\rho \right)$ 
($i=1,\dots,k$, $\rho \in {\cal M}_N$), where the linear operators ${\cal L}_i$ are 
Hermitian, positive, and fulfill the identity $\sum_{i=1}^{k}{\cal L}_{i}=1$.

Now the dynamics takes place in the convex body of all density matrices
${\cal M}_N$. The space of mixed states ${\cal M}_N$ has $N^2-1$ real
dimensions in contrast to the $\linebreak(2N-2)-$dimensional space of pure states
${\cal P}_N$. For $N=2$ its is just the $3$--dimensional {\sl Bloch ball}, i.e.,
the volume bounded by the Bloch sphere.

The special class of QIFSs is a class of {\sl homogenous} QIFSs introduced in 
more general setting by one of the authors \cite{Slo02}. A QIFS is called {\sl 
homogenous} if both $p_i$ and $G_i \cdot p_i$ are affine maps for
$i=1,\dots,k$. The mixed states QIFS being a generalization of a pure state QIFS 
and defined by formulas (\ref{mapsmix}) and (\ref{probabilitiesmix}) is 
homogenous if $W_i=V_i$ for $i=1,\dots,k$. Interesting examples of such
systems acting on the Bloch sphere  where recently analyzed by Jadczyk
and {\"O}berg \cite{JO02}. For a homogenous QIFS $p_i$ and $G_i$ 
may be interpreted in terms of a discrete measurement process as the probability 
that the measurement outcome is $i$, and the state of the system after the 
measurement if the result was actually $i$, respectively.

A homogenous QIFS generates not only the Markov operator $P$ acting in the space 
of probability measures on ${\cal M}_N$, but also the linear, trace-preserving, 
and positive operator $\Lambda : {\cal M}_N \to {\cal M}_N$ defined by 
\begin{equation}
\Lambda (\rho) =\sum_{i=1}^k p_i(\rho) G_i(\rho) = \sum_{i=1}^k V_i\rho
V_i^{\dagger}
\label{rhoqifs}
\end{equation}
for $\rho \in {\cal M}_N$.

A mixed state $\widetilde{\rho}$ is $\Lambda-$invariant if and only if it is the
{\sl barycenter} of some $P-$invariant measure $\widetilde{\mu}$, i.e.,
\begin{equation}
\widetilde{\rho }=\int_{\cal M_{N}}\rho d \widetilde{\mu }\left( \rho \right) \ ,
\label{invmixa}
\end{equation}
see Ref.~\cite{Slo02}.
\medskip

\example  $\Omega={\cal M}_N$, $k=2$, $p_1=p_2=1/2$,
$G_1(\rho)=U_1\rho U_1^{\dagger}$ and $G_2(\rho)=U_2\rho U_2^{\dagger}$. 
This is just Example~6 in other casting; the normalized identity matrix,
$\rho_* = {\mathbbm 1}/N$ is $\Lambda-$invariant irrespectively of the form of
unitary operators $U_i, i=1,2$. Note that $\widetilde{\rho}=\rho_*$ may be 
represented as Eq.~(\ref{invmixa}), where the measure $\widetilde{\mu}$, 
uniformly spread over ${\cal P}_N$ (the {\sl Fubini-Study measure}),
is $P$ invariant.
\medskip

To define hyperbolic QIFSs one needs to specify a distance in the space of mixed 
quantum states. There exist several different metrics in ${\cal M}_N$, which may 
be applicable (see e.g. Ref.~\cite{PS96,ZS01} and references therein). The standard 
distances: the {\sl Hilbert-Schmidt distance}
\begin{equation}
D_{HS}(\rho_1,\rho_2) = \sqrt{{\rm tr}[
(\rho_1-\rho_2)^2]}
\text{ ,}
\label{HS}
\end{equation}
the {\sl trace distance}
\begin{equation}
D_{\mbox{\scriptsize{tr}}}(\rho_1,\rho_2) = {\rm tr}\sqrt{(\rho_1-
\rho_2)^2} 
= \mid \mid \rho{}_{1}-\rho{}_{2}\mid \mid
\!{}_{\raisebox{-1mm}{\scriptsize{tr}}}
\text{ ,}
\label{trano}
\end{equation}
and the {\sl Bures distance} \cite{bu69}
\begin{equation}
D_{Bures}(\rho_1,\rho_2)=\sqrt{ 2\bigl\{1-{\rm tr
[(\rho _1^{1/2}\rho_2\rho_1^{1/2})^{1/2}]\bigr\} }}
\label{Bures}
\end{equation}
the latter based on the idea of purification of mixed quantum states
\cite{ul76,hu92}, are mutually bounded \cite{Ul2}. They generate the same 
natural topology in ${\cal M}_N$. Having endowed the space of mixed state with a 
metric, we may formulate immediate conclusion from the theorem on hyperbolic 
IFSs. We define a hyperbolic QIFS as in the previous section, and the
following proposition holds. 

\medskip

\proposition {\sl If a QIFS (\ref{QIFSb}) is homogenous and hyperbolic
(that is, the quantum maps $G_i$ are contractions with respect to one of
the standard distances in ${\cal M}_N$, $p_i$ are H\"{o}lder continuous
and positive), then the associated Markov operator $P$ possesses a
unique invariant measure $\widetilde{\mu}$. This invariant measure
determines a unique $\Lambda- $invariant mixed state $\widetilde{\rho}
\in {\cal M}_N$ given by Eq.~(\ref{invmixa}).}

\medskip
Note that for a homogenous hyperbolic QIFS, the sequence $\Lambda ^n(\rho_0)$ 
tends in the limit $n \to \infty$ to a unique invariant state $\widetilde{\rho}$ 
irrespectively of the choice of an initial state $\rho_0$ \cite{Slo02}.

\example 
$\Omega={\cal M}_N$, $k=2$, $p_1=p_2=1/2$,
$G_1(\rho)=(\rho+2\rho_1)/3$ and $G_2(\rho)=(\rho+2\rho_2)/3$, where we
choose the both projectors $\rho_1=|1\rangle\langle 1|$ and
$\rho_1=|2\rangle\langle 2|$ to be orthogonal. Since both homotheties $G_i$ are
contractions (with the Lipschitz constants $1/3$) this QIFS is hyperbolic
and a unique invariant measure $\widetilde{\mu}$ exists. In analogy
with the IFS discussed in Example 1 we see that the support of $\widetilde{\mu}$ 
covers the Cantor set at the line joining both projectors $\rho_1$ and $\rho_2$.
However, this is nothing but a rather sophisticated representation
of the maximally mixed two-level state $\rho_* = (\rho_1+\rho_2)/2$, which 
follows from the symmetry of the Cantor set and may be formally verified by 
performing the integration prescribed by Eq.~(\ref{invmixa}).

\subsection{Completely positive maps and unitary QIFSs}

From the mathematical point of view it may be sufficient to require
that the map $\Lambda$ is {\sl positive}, that is, it transforms a positive
operator into another positive operator. From the physical point
of view it is desirable to require a stronger condition of complete
positivity related to a possible coupling of the quantum system under
consideration with an environment. A map $\Lambda$ is {\sl completely positive}
({\sl CP-map}), if the extended map $\Lambda \otimes {\mathbbm 1}$ is positive
for any extension of the initial Hilbert space, ${\cal H}_N \to {\cal
H}_N \otimes {\cal H}_E$, which describes coupling to the environment
\cite{Kr71,Li75}.

It is well known that each trace preserving CP-map $\Lambda$ (sometimes called 
{\sl quantum channel}), can be represented (non uniquely) in the following {\sl 
Stinespring-Kraus form}
\begin{equation}
\rho'= \Lambda_K(\rho)= \sum_{j=1}^{k} V_j \rho
V_j^{\dagger}, \quad {\rm with} \quad  \sum_{j=1}^{k} V_j^{\dagger} V_j
= {\mathbbm 1} \text{ ,}
\label{krauss}
\end{equation}
where linear operators $V_j$ ($j=1,\dots,k$) are called {\sl Kraus operators} 
\cite{Kr71,Ch75}. For any quantum channel acting in an $N-$dimensional
Hilbert space the number of operators $k$ needs not exceed $N^2$
\cite{Kr83}. Each quantum channel can be treated (but not necessarily
uniquely) as a pure or mixed states homogenous QIFS. Conversely, for
each homogenous QIFS, formula (\ref{rhoqifs}) defines a quantum channel.

If, additionally, $\sum_{j=1}^{k} V_j V_j^{\dagger}= {\mathbbm 1}$
holds, then $\Lambda({\mathbbm 1}/N)={\mathbbm 1}/N$, and the map
$\Lambda$ is called {\sl unital}. It is the case if all Kraus 
operators are normal, $V_j V_j^{\dagger}=V_j^{\dagger} V_j$ ($j=1,\dots,k$), 
however, this condition is not necessary. A unital trace preserving
CP-map is called {\it bistochastic}. An example of a
bistochastic channel is given by {\it random external fields}
\cite{Al87} defined by
\begin{equation}
\rho' = \Lambda_{U}(\rho)= \sum_{i=1}^k p_i ~ U_i \rho
U_i^\dagger
\text{ ,}
\label{refiel}
\end{equation}
where $U_{i}$, $i=1,2,\dots,k$ are {\it unitary} operators and the vector of
non-negative probabilities is normalized, i.e., \linebreak 
$\sum_{i=1}^{k}p_{i}=1$.
The Stinespring-Kraus form (\ref{krauss}) can be reproduced setting $V_i = 
\sqrt{p_i} U_i$. Note that the random external field (\ref{refiel}) may be 
regarded as a homogenous QIFS of the first kind (with constant probabilities) 
with $k$ unitary maps $G_i(\rho)= U_i \rho U_i^\dagger$ ($i=1,\dots,k$). In 
particular, Example~7 belongs to this class. In the sequel such QIFSs will be 
called {\sl unitary}. For a unitary QIFS not only $\rho_*$ is an invariant state 
of $\Lambda_U$, but also the measure $\delta_{\rho_*}$ is invariant for the 
operator $P_U$ induced by this QIFS.

Although a unitary QIFS consists of isometries, the operator $\Lambda_U$ needs 
not preserve the standard distances between any two mixed states. For the 
Hilbert-Schmidt metric we have
\begin{equation}
D_{HS}\bigl(\Lambda_U(\rho_1),\Lambda_U(\rho_2) \bigr)\le
D_{HS}\bigl( \rho_1, \rho_2 \bigr)
\text{ .}
\label{contHS}
\end{equation}
In fact this statement is true for any bistochastic channels as shown by
Uhlmann \cite{Ul71}, but it is false for arbitrary CP maps, since the
Hilbert-Schmidt metric is not monotone \cite{Oz00}.  On the other hand,
$\Lambda_U$ is a contraction for the Bures distance (Riemannian) and the
trace distance (not Riemannian), which are monotone and do not grow
under the action of any CP map \cite{Ru94,PS96}.  Choosing for $\rho_2$
the maximally mixed state $\rho_* = {\mathbbm 1}/N$, which is invariant
with respect to $\Lambda_U$ for any unitary QIFS, we see in particular
that the distance of any state $\rho_1$ to $\rho_*$ does not increase in
time.  Similarly, the von Neumann entropy given by  $H(\rho)={\rm
tr}(\rho\ln \rho )$ for $\rho\in{\cal M}_N$ does not decrease during the
time evolution (\ref{refiel}). On the other hand, the inequality in
Eq.~(\ref{contHS}) is weak, and in some cases the distance may remain
constant. The question, under which conditions this inequality is
strong, is related to the problem, for which unitary QIFSs the maximally
mixed state $\rho_*$ is a unique invariant state of $\Lambda_U$. This is
not the case, if all operators $U_i$ commute, since then all density
matrices diagonal in the eigenbase of $U_i$ are invariant. Such a
situation may occur also in subspaces of smaller dimension. To describe
such a case we shall call unitary matrices of the same size {\sl common
block--diagonal}, if they are block-diagonal in the same basis and with
the same blocks. The uniqueness of the invariant state of a unitary QIFS
is then characterized by the following proposition, the proof of which
is provided in the appendix A.

\medskip

\proposition {\sl Let us assume that all probabilities $p_i$ ($i=1,\dots,k$) are 
strictly positive. Then the maximally mixed state $\rho_*$ is not a unique 
invariant state for the operator $\Lambda_U$ if and only if unitary operators 
$U_i$ ($i=1,\dots,k$) are common block-diagonal.}
\label{blodia}
\medskip

It follows from the proof of this proposition that in this case there exists 
$\rho \neq \rho_*$ such that $\delta_{\rho}$ is an invariant measure for the 
operator $P_U$ induced by the QIFS.

To show an application of Proposition~2 consider a two level quantum
system, called {\sl qubit}, which may be used to carry a piece of quantum 
information. Let us assume it is subjected to a random noise, described
by the following map:
\begin{equation} 
\rho \to \rho' = \Lambda_U(\rho)=(1-p)\rho + \frac{p}{3} \bigl[
\sigma_1 \rho \sigma_1 + \sigma_2 \rho \sigma_2 + \sigma_3 \rho \sigma_3
\Bigr]
\text{ .}
\label{depol}
\end{equation}

This bistochastic map, defined by
the unitary Pauli matrices $\sigma_i$,  is called {\sl depolarizing
quantum channel}  \cite{Pres}, and the parameter $p$ plays the role of
the probability of error. This map transforms any vector inside the
Bloch ball toward the center, so the length of the polarization vector
decreases. In formalism of QIFSs this quantum channel is equivalent to
the following example.

\medskip

\example  $\Omega={\cal P}_2$, $k=4$, $U_1={\mathbbm 1}$, $U_2=\sigma_1$,
$U_3=\sigma_2$, $U_4=\sigma_3$, $p_1=1-p$ and $p_2=p_3=p_4=p>0$. Since the Pauli 
matrices are not common block-diagonal, the maximally mixed state $\rho_*$ is a 
unique invariant state of the CP map (\ref{depol}) associated with this 
unitary QIFS.

\medskip

To introduce an example of QIFS arising from atomic physics, consider a two level 
atom in a constant magnetic field $B_z$ subjected to a sequence of resonant 
pulses of electromagnetic wave. The length of each wave pulse is equal to its 
period $T$ and it interacts with the atom by the periodic Hamiltonian 
$V(t)=V(t+T)$. Let us assume that each pulse occurs randomly with probability 
$p$. Thus, the evolution operator transforms any initial pure state by
the operator
\begin{equation}
U_1=\exp(-iH_0T/\hbar)
\label{atomic1}
\end{equation}
in the absence of the pulse, or by the operator
\begin{equation}
U_2={\hat C}\exp \left[ -\frac{i}{\hbar}\left( H_0 T+\!\! \int_0^T \!\!\!\!
V(t)\mbox{d$t$}\right) \right]
\label{atomic2}
\end{equation}
in the presence of the pulse. The unperturbed Hamiltonian $H_0$ is proportional
to  $B_z J_z$ ($J_z$ is $z$ component of the angular momentum operator) and
${\hat C}$ denotes the chronological operator. Thus, this random system may be
described by the following QIFS.

\medskip

\example  $\Omega={\cal P}_2$, $k=2$, $p_1=1-p$ and $p_2=p$,
the Floquet operators $U_1$ (\ref{atomic1}) and $U_2$ (\ref{atomic2})
as specified above. The maximally mixed state $\rho_* = {\mathbbm 1}/2$, 
corresponding to the center of the Bloch ball, is the invariant state of the 
Markov operator given by Eq.~(\ref{refiel}). For the case of a generic
perturbation $V$, the matrices $U_1$ and $U_2$ are not common
block-diagonal, and so $\rho_*$ is the unique invariant state for
operator (\ref{refiel}) related to the QIFS.

\medskip

The QIFSs arise in a natural way if considering a quantum system acting on 
${\cal H}_N$ coupled with an {\sl ancilla}: a state 
in an auxiliary $m$-dimensional Hilbert space ${\cal H}_m$, which 
describes the environment. Initially, the composite state describing the system 
and the environment is in the product form, $\sigma=\rho_A \otimes \rho^{B}_*$, 
where $\rho^{B}_*={\mathbbm 1}_m/m$ is the maximally mixed state,  
but the global unitary evolution couples two subsystems together. A unitary 
matrix $U$ of size $Nm$ acting on the tensor space ${\cal H}_N \otimes {\cal
H}_m$ may be represented in its Schmidt decomposition form as $U=
\sum_{i=1}^K \sqrt{q_i} V^{A}_i \otimes V^{B}_i$, where the number of terms is
determined by the size of the smaller space, $K = \min \{N^2,m^2\}$; the
operators $V^{A}_i$ and $V^{B}_i$ act on ${\cal H}_N$ and ${\cal H}_k$
respectively, and the Schmidt coefficients are normalized as
$\sum_{i=1}^K q_i=1$. Restricting our attention to the system $A$ one
needs to trace out the variables of the environment $B$ which leads to
the following quantum channel (and to the respective homogenous QIFS):
\begin{equation}
\rho_A'= \Lambda(\rho_A) = {\rm tr}_B (U \sigma U^{\dagger})=
\sum_{i=1}^K q_i V^{A}_i \rho_A  V^{A}_i{}^\dagger
\text{ .}
\label{refie2}
\end{equation}
Since for $\rho^A_{*}={\mathbbm 1}_N/N$ we have $\Lambda(\rho^{A}_*)=\rm
tr_B(U(\rho^{A}_* \otimes \rho^{B}_*)U^{\dagger})=\rho^{A}_*$, the
CP-map $\Lambda$ is bistochastic.

\section{Quantum--Classical correspondence}

To investigate various aspects of the semiclassical limit of the quantum
theory it is interesting to compare a given discrete classical dynamical system 
generated by $f: \Omega \to \Omega$ with a family of the corresponding quantum 
maps, usually defined as 
$F_N: {\cal H}_N \to {\cal H}_N$ with an integer $N$. 
Several alternative methods of quantization of classical maps in compact phase 
space have
been applied to construct quantum maps corresponding to baker map on the
torus \cite{BV87,Sa90}, Arnold cat map \cite{Ke91} and other
automorphisms on the torus \cite{BEG96}, periodically kicked top
\cite{KSH87} and baker map on the sphere \cite{POZ99}.

To specify in which manner the classical and the quantum maps are
related, it is convenient to introduce a set of coherent states
$|y\rangle \in {\cal H}_N$, indexed by classical points $y$ of the phase
space $\Omega$. (For more properties of coherent states and a general
definition consult the book of Perelomov \cite{Pe86}.) They satisfy the
resolution of identity formula: $\int_{\Omega} |y\rangle \langle y| dy
={\mathbbm 1}$, and allow us to represent any state $\rho$ by its Husimi
representation, $H(y)=\langle y |\rho| y \rangle$i ($y\in\Omega$).
Quantization of a classical map $f$, which leads to a family of quantum
maps $F_N$ is called {\sl regular}, if for almost all classical points
$x$ the classical and the quantum images are connected in the sense that
the normalized Husimi distribution of the state $F_N|y\rangle$
integrated over a finite vicinity of the point $f(y)$ tends to unity in
the limit $N\to \infty$ \cite{SZ94}. Another method of linking a
classical map with a family of quantum maps is based on the {\sl Egorov
property}, which relates the classical and the quantum expectation
values \cite{Om97,BE98}.

In a similar way we may construct QIFSs related to certain
classical IFSs. More precisely, a sequence of pure states QIFS ${\cal F}_N = \{ 
{\cal P}_N; F_{i,N},p_{i,N} : i=1,\dots,k\}$ induced by two sets of linear maps 
$V_{i,N}, W_{i,N}: {\cal H}_{N} \rightarrow {\cal H}_{N}$ ($i=1,\ldots ,k$) 
(see (\ref{maps}) and (\ref{probabilities})) is a 
{\sl quantization} of a classical IFS ${\cal F}_{\rm Cl} = \{\Omega;F_i,p_i:1,\dots,k\}$, 
when:

\begin{itemize}
\item the functions $F_{i,N}$ are quantum maps obtained by quantization of the 
classical maps $f_i$;
\item the probabilities $p_{i,N}$ computed at coherent states $|y\rangle$  fulfill
\begin{equation}
p_{i,N} (|y\rangle \langle y |) = 
\left\| W_{i,N}\left( \left| y\right\rangle \right) \right\| ^{2} 
\stackrel{ N \to \infty}
{\longrightarrow} p_i(y) \quad {\text {\ for }} \quad y \in \Omega {\text{\ 
and } } i=1,...,k
\text{ .}
\label{probclaq}
\end{equation}
\end{itemize}

To illustrate the procedure let consider random rotations on the sphere,
performed along $x$ or $z$ axis. This special case of Example 4 may
be easily quantized with the help of the components $J_i$ ($i=x,y,z$) of the 
angular momentum operator $J$, satisfying the standard commutation relations,
$[J_i,J_j]=\epsilon_{ijk}J_k$. The size of the Hilbert space is
determined by the quantum number $j$ as $N=2j+1$.

\medskip

\example $k=2$, random rotations are given as the following

a) classical, ${\cal F}_{\rm Cl} = \{\Omega=S^2,$ $f_1=R_z(\theta_1),$
$f_2=R_x(\theta_2),$ $p_1=p_2=1/2 \}$. The Lebesgue measure on the sphere is an 
invariant measure of this IFS.

b) quantum, ${\cal F}_N=\{\Omega={\cal P}_N,$ $F_1=\exp(i\theta_1 J_z),$
$F_2=\exp(i \theta_2 J_x),$  $p_1=p_2=1/2 \}$. Since both unitary
operators are not common-block diagonal, due to Proposition 2, the maximally 
mixed state $\rho_*$ is a unique invariant state for operator
(\ref{rhoqifs}) to the QIFS ${\cal F}_N$.

\medskip

A quantization of an IFS of the second kind is given by the following
modification of the previous example.

\medskip

\example  $k=2$, random rotations on the sphere with varying
probabilities depending on the latitude $\theta$ computed with respect
to the $z$ axis. 

The spaces and the functions are as in Example~11, but

a) classical IFS ${\cal F}_{\rm Cl}$: $p_1=(1+\cos \theta)/2$ and
  $p_2=(1-\cos \theta)/2$;

b) quantum IFS  ${\cal F}_N$: $p_1=1/2 +\langle J_z \rangle /2j $  and
  $p_2=1/2 -\langle J_z \rangle /2j $ with $N=2j+1$.
  Interestingly, this modification
  influences the number of invariant states of the IFS. Since $p_2$
  vanish at the north pole, $\theta=0$, of the classical sphere $S^2$,
  this point is invariant with respect to ${\cal F}_{\rm Cl}$.
  Similarly, the corresponding quantum state $|j,j\rangle$  localized at
  the pole is invariant with respect to the QIFS ${\cal F}_N$.
  
\medskip

The above examples of unitary QIFS dealt with simple regular maps ---
rotations on the sphere. However, an IFS may also be constructed out of
nonlinear maps, which may lead to deterministic chaotic dynamics. For
instance, one may consider the map describing periodically kicked top. It
consists of a linear rotation with respect to $x$ axis by angle
$\alpha$ and a nonlinear rotation with respect to $z$ axis by an angle
depending on the $z$ component. In a compact notation the classical top
reads, $T_{\rm Cl}(\alpha,\beta):=R_z(z\beta)R_x(\alpha)$, while its
quantum counterpart, acting in the $N=2j+1-$dimensional Hilbert space
can be defined by $T_{Q}(\alpha,\beta):=\exp(-i\beta J_z^2/2j)
\exp(-i\alpha J_x)$ \cite{KSH87}. This quantum map becomes one of the
important toy model often studied in research on quantum chaos
\cite{Ha00}. A certain modification of this model, in which the kicking
strength parameter $\beta$ was chosen randomly out of two values, was
proposed and investigated by Scharf and Sundaram \cite{SS94}. This
random system may be put into the QIFSs formalism.

\medskip

\example Randomly kicked top.

a) classical, ${\cal F}_{\rm Cl}=\{ \Omega=S^2,$ $f_1=T_{\rm Cl}
(\alpha,\beta),$ $f_2=T_{\rm Cl} (\alpha,\beta+\Delta),$ $p_1=p_2=1/2
\}$.

b) quantum, ${\cal F}_N=\{ \Omega={\cal P}_N$, $F_1=T_{Q}
(\alpha,\beta)$, $f_2=T_{Q}(\alpha,\beta+\Delta)$, $p_1=p_2=1/2 \}$.
For $\alpha>0$ and a positive $\Delta$ both unitary operators are not
block-diagonal, so the maximally mixed state $\rho_*$ is a unique
invariant state for operator (\ref{rhoqifs}) related to the unitary
QIFS.  Our numerical results obtained for $\alpha=\pi/4$, $\beta=2$ and
$\Delta=0.05$ suggest that the trajectory of any pure coherent state
converges to the equilibrium exponentially fast.

\medskip

To discuss a quantum analogue of an IFS with a fractal invariant measure
consider the classical IFS presented in Example 3. The classical phase
space $\Omega$ is equivalent to the torus. For pedagogical purpose, let
us rename both variables $x,y$ into $q,p$, represented canonically
coupled position and momentum. We shall work in $N=3L-$dimensional
Hilbert space. Let $|j\rangle_q$ with $j=1,...,N$ be eigenstates of the
position operator, and similarly $|l\rangle_p$ with $l=1,...,N$ be the
eigenstates of the momentum operator. Both bases are related by
$|l\rangle_p = \sum_{j=1}^N W_{lj} |j\rangle_q$, where the matrix $W$ is
the $N$ point discrete Fourier transformation with $W_{lj} =
(1/\sqrt{N}) e^{-2\pi ilj/N} $. The classical map $f_1$ in
Eq.~(\ref{clascant}), representing a three--fold contraction in the $x$
direction, corresponds to the transformation $G_1$ of the density
operator given by

\begin{equation}
G_1(\rho ) = \sum_{i,j=1}^L |i\rangle_q
\left( \sum_{m,n=0}^2 \langle 3i+m|_q\rho |3j+n\rangle_q \right)
\langle j|_q
\text{ .}
\end{equation}

In a similar way, the quantum map $G_2$ corresponding to $f_2$ is defined by
\begin{equation}
G_2(\rho ) = \sum_{i,j=2L+1}^{3L} |i\rangle_q
\left( \sum_{m,n=0}^2 \langle 3i+m|_q\rho |3j+n\rangle_q \right)
\langle j|_q
\text{ .}
\end{equation}

The maps $G_3$ and $G_4$ are obtained in analogous way like $G_1$ and $G_2$,
using the eigenstates of momentum operator $|k\rangle_p$,
\begin{eqnarray}
G_3(\rho ) = \sum_{k,l=1}^L |k\rangle_p
\left( \sum_{m,n=0}^2 \langle 3k+m|_p\rho
|3l+n\rangle_p \right) \langle l|_p
\text{ ,} \\
G_4(\rho ) = \! \! \sum_{k,l=2L+1}^{3L} |k\rangle_p
\left( \sum_{m,n=0}^2 \langle 3k+m|_p\rho
|3l+n\rangle_p \right) \langle l|_p \text{ .}
\end{eqnarray}
The random system defined below may be considered as a QIFS related to the
IFS introduced in Example~3.

\medskip

\example Quantum tartan specified by the following QIFS:
 ${\cal F}_N=
\{\Omega={\cal P}_N$, $k=4$, $G_1,G_2,G_3,G_4$;
$p_1=p_2=p_3=p_4=1/4 \}.$

\medskip

An invariant states for the maps $\Lambda$ induced by this QIFS are 
illustrated in Fig.~1 for $N=3^4$, $N=3^5$ and $N=3^6$. Invariant quantum state 
$\rho_{*}$ is shown in the generalized Husimi representation

\begin{equation}
H_{\rho}(p,q)= \frac{1}{2\pi} \frac{ \langle
q,p|\rho|q,p\rangle }{ \langle q,p |q,p \rangle}
\text{ ,}
\label{hustor}
\end{equation}
based on the set of coherent states on the torus $|q,p\rangle=
Y^{Np-N/2} X^{Nq-N/2} |\kappa\rangle$.  The reference state $|\kappa\rangle$
is chosen as an arbitrary state localized in $(1/2,1/2)$
\begin{equation}
\langle n | \kappa \rangle = (2/N)^{-1/4}e^{-\pi (n-N/2)^2/N-i\pi n}
\text{ ,}
\label{gstate}
\end{equation}
while $X$ denotes the operators of shift in position $X|j\rangle =
|j+1\rangle$, with an identification $|j+N\rangle = |j\rangle$ for
$j=1,\dots{},N$. Similarly $Y$ shifts the momentum eigenstates,
$Y|l\rangle = |l+1\rangle$ and $|l+N\rangle = |l\rangle$ for
$l=1,\dots{},N$. The quantum state $|q,p\rangle$ is well localized in
the vicinity of the classical point $(q,p)$ on the torus \cite{Va99}.
This representation of quantum states corresponding to the classical
system on the torus was used in the analysis of an irreversible quantum
baker map \cite{LP02}.

The larger value of $N$, the finer structure of the invariant state
$\rho_*$ is visible in the phase space. In the semiclassical limit
$N\to\infty$, (which means $\hbar \to 0$) the invariant state $\rho_{*}$
tends to be localized at the fractal support of the invariant measure of
the classical IFS, shown for comparison in Fig.~1c. Strictly speaking,
for any finite $N$, the Husimi distribution of the quantum state
$\rho_*$ does not posses fractal character, since self-similarity has to
terminate at the length scale comparable with  $\sqrt{\hbar}$. In other
words, quantum effects are responsible for smearing out the fractal
structure of the classical invariant measure. However, the classical
fractal structures may be approximated with an arbitrary accuracy by
quantum objects in the semiclassical limit \cite{WBZ00}.

\section{Closing Remarks}

Classical iterated function systems display several interesting
mathematical properties and may be applied in various problems from
different branches of physics. In this work we have generalized the
formalism of IFSs introducing the concept of QIFSs. Quantum iterated
function systems may be defined in the space of pure states on a
finite dimensional Hilbert space ${\cal H}_N$, or more generally, in the
space of density operators acting on ${\cal H}_N$. As their
classical analogues, QIFSs allow a certain degree of stochasticity, in
the sense that at each step of time evolution the choice of one of the 
prescribed quantum maps is random.

This formalism is useful to describe several problems of quantum
mechanics, including non-unitary dynamics, processes of decoherence and
quantum measurements. In fact, the large class of quantum channels,
called random external fields may serve directly as
examples of a QIFS. Furthermore, for several classical IFSs one may
construct the corresponding QIFSs and analyze the similarities and
differences between them. As shown in the last example, one may focus on
the fractal properties of invariant measures of some classical IFSs and
study their quantum counterpart. Thus the concept of QIFS allows one to
investigate the semiclassical limit of random quantum systems.

K.{\.Z}. is grateful to E. Ott for fruitful discussions and hospitality at
the University of Maryland, where this work was initiated. We are also
thankful to R. Alicki, M. Fannes, M. Ku{\'s} and P. Pako{\'n}ski for
helpful remarks. Financial support by Polski Komitet Bada{\'n}
Naukowych under grant number 2P03B~072~19 is gratefully acknowledged.

\appendix

\section{Proof of Proposition 3}
We start from the following lemma:

\medskip

\textbf{Lemma 1.} Let $U=\left(  U_{nm}\right)  _{n,m=1,\ldots,N}$ be an
$N-$dimensional unitary matrix. Assume that there exist two non-empty sets of
indices $A$ and $B$ such that: $A\cup B=I:=\left\{  1,\ldots,N\right\}  $ and
$A\cap B=\emptyset$. Then, $U_{nm}=0$ for $n\in A$ and $m\in B$, implies
$U_{nm}=0$ for $n\in B$ and $m\in A$. 

\medskip

\textbf{Proof of the lemma:} We compute the number of elements of the set $A$:
\begin{align*}
\left|  A\right|    & =\sum_{n\in A}\sum_{m\in I}\left|  U_{nm}\right|
^{2} \nonumber\\
& =\sum_{n\in A}\sum_{m\in A}\left|  U_{nm}\right|  ^{2}+\sum_{n\in A}%
\sum_{m\in B}\left|  U_{nm}\right|  ^{2} \nonumber \\
& =\sum_{n\in A}\sum_{m\in A}\left|  U_{nm}\right|  ^{2} \nonumber \\
& =\sum_{n\in I}\sum_{m\in A}\left|  U_{nm}\right|  ^{2}-\sum_{n\in B}%
\sum_{m\in A}\left|  U_{nm}\right|  ^{2} \nonumber \\
& =\left|  A\right|  -\sum_{n\in B}\sum_{m\in A}\left|  U_{nm}\right|
^{2}\text{ ,}
\end{align*}
and so $\sum_{n\in B}\sum_{m\in A}\left|  U_{nm}\right|  ^{2}=0$, as required.

\medskip
\medskip

Now we turn to the proof of Proposition 3. 

$\Rightarrow)$\textbf{ }Let $U_{i}$ ($i=1,,\ldots,k$) be block-diagonal in the
common base, and let dimension of the blocks be 
$\alpha_{1},\ldots,\alpha_{L}$,
 where $\sum_{j=1}^{L}\alpha_{j}=N$.
 Define a diagonal density matrix as a direct sum 
\begin{equation}
\rho:=\bigoplus_{j=1}^{L}\frac{\sigma_{j}}{\alpha_{j}}
{\mathbbm 1}_{\alpha_{j}}\text{ ,}
\end{equation}
where $\sum_{j=1}^{L}\sigma_{j}=1$. Then, $U_{i}\rho U_{i}^{\dagger}=\rho$ for
every $i=1,\ldots,k$. Hence $\rho$ is $\Lambda_{U}-$invariant and
$\delta_{\rho}$ is a $P_{U}-$invariant measure on ${\cal P}_N$ for an
arbitrary choice of $\left(  \sigma_{j}\right) _{j=1,\ldots,L}$.

\medskip

$\Leftarrow)$\textbf{ }Let $\rho$ be an invariant state for $\Lambda_{U}$
such that $\rho \ne \rho_*$.
Then $\rho$ can be written in the form
\begin{equation}
\rho=\sum_{n=1}^{N}\sigma_{n}|\Psi_{n}\rangle\langle\Psi_{n}|\text{ ,}
\end{equation}
where $\left|  \Psi_{m}\right\rangle \in\mathcal{P}_{N}$, $\langle\Psi
_{n}|\Psi_{m}\rangle=\delta_{nm}$ ($n,m=1,\dots,N$), and $\sigma_{1}\leq
\sigma_{2}\leq\dots\leq\sigma_{N}$; $\sigma_1 \le 1/N$.
For $\gamma \in [0,1]$ the density operator 
$\rho^{\prime}=\gamma\rho+(1-\gamma)\rho_{\ast}
=\sum_{n=1}^{N}\sigma_{n}^{\prime}|\Psi_{n} \rangle\langle\Psi_{n}|$,
 where $\sigma_{n}^{\prime}=\gamma\sigma_{n}+\left(
1-\gamma\right)  N^{-1}$ ($n=1,\ldots,N$) is also an invariant state for
$\Lambda_{U}$. Put $\gamma:=1/(1-\sigma_{1}N)$. This choice implies
$\sigma_{1}^{\prime}=0$ and $\sum_{n=1}^{N}\sigma_{n}^{\prime}=1$. Assume that
$\sigma_{n}^{\prime}=0$ for $n=1,\ldots,n^{\prime}$ and $\sigma_{n}^{\prime
}>0$ for $n=n^{\prime}+1,\ldots,N$, where $n^{\prime}\geq1$. The equation
$\Lambda_{U}(\rho^{\prime})=\rho^{\prime}$ can be rewritten in the form
\begin{equation}
\sigma_{n}^{\prime}=\sum_{i=1}^{k}p_{i}\sum_{m=1}^{N}|(U_{i})_{nm}|^{2}
\sigma_{m}^{\prime}\text{ ,}%
\end{equation}
where $(U_{i})_{nm}$ ($n,m=1,\dots,N$) are the elements of matrices
$U_{i}$ ($i=1,\ldots,k$) in the basis $\left(  \left|  \Psi_{n}\right\rangle
\right)_{n=1,\ldots,N}$.

For $n=1,\dots,n^{\prime}$ we get
\begin{equation}
0=\sum_{i=1}^{k}p_{i}\sum_{m=n^{\prime}+1}^{N}|(U_{i})_{nm}|^{2}\sigma
_{m}^{\prime}\text{ .}
\end{equation}
Hence $(U_{i})_{nm}=0$ for $n=1,\dots,n^{\prime}$ and $m=n^{\prime}+1,\dots
,N$. Using Lemma 1, we deduce that $(U_{i})_{nm}=0$ for $n=n^{\prime}%
+1,\dots,N$ and $m=1,\dots,n^{\prime}$. Thus $U_{i}$ ($i=1,\ldots,k$) are
common block-diagonal.


\end{document}